\documentclass{ws-ijmpd}
\usepackage[super,compress]{cite}
\usepackage{epsfig}
\usepackage{graphicx,amssymb}
\usepackage{bm}
\usepackage{colordvi}
\usepackage{color}
\usepackage{changes}

\begin{document}
\markboth{E. Piedipalumbo, E. Della Moglie, R. Cianci}
{Upadted high redshift $f(T)$ cosmography}

%
\catchline{}{}{}{}{}
%

\title{Updated F(T) gravity constraints from high redshift cosmography }

\author{Ester Piedipalumbo}

\address{Dipartimento di Fisica, Universit\`{a} di Napoli
Federico II, Compl.
Univ. Monte S. Angelo, 80126 Napoli, Italy\\
 I.N.F.N., Sez. di Napoli, Complesso Universitario di Monte
Sant' Angelo, Edificio G, via Cinthia, 80126 Napoli, Italy\\
ester@na.infn.it}

\author{Enrica Della Moglie }

\address{DIME Sezione Metodi e Modelli Matematici,  Universit\`{a} di Genova,\\\,\, P.le J.F. Kennedy, I 16129  Genova, Italy\\
dellamoglie@dime.unige.it}

\author{Roberto Cianci}

\address{DIME Sezione Metodi e Modelli Matematici,   Universit\`{a} di Genova,\\\,\, P.le J.F. Kennedy, I 16129  Genova,Italy\\
cianci@dime.unige.it}

\maketitle

\begin{history}
\received{Day Month Year}
\revised{Day Month Year}
\end{history}

\begin{abstract}
In the last dozen years a wide and variegated mass of observational data revealed that the universe is now
expanding at an accelerated rate. In the absence of a well-based theory to interpret the observations,
cosmography provides  information about the evolution of the Universe from measured distances, only assuming that the geometry of the can be described by the Friedmann-Lemaitre-Robertson -Walker metric. We perform a  high-redshift analysis allows us to put constraints on the cosmographic parameters up to the fifth order, thus inducing indirect constraints on any gravity theory. Here we are interested in the so called teleparallel gravity  theory, $f(T)$. Actually we use the analytical expressions of the present day values of $f(T)$ and its derivatives as  functions of the cosmographic parameters to map the {\it cosmography region of confidences}  into confidence ranges for $f(T)$ and its derivative. Moreover, we show how these can be used to test some teleparallel gravity  models without solving the dynamical equations. Our analysis is based on the Union2 Type Ia Supernovae (SNIa) data set,  a set of 28  measurements of the Hubble parameter,  the Hubble diagram constructed from some Gamma Ray Bursts (GRB) luminosity distance indicators, and gaussian priors on the distance from the Baryon Acoustic Oscillations (BAO), and the Hubble constant $h$. To perform our statistical analysis and to explore the probability distributions of the cosmographic parameters we use the Markov Chain Monte Carlo Method (MCMC).  \end{abstract}


\ccode{PACS numbers:98.80.-k; 95.36.+x,  04.50.Kd}


\section{Introduction }	

Over the last dozen years a wide and variegated mass of observations revealed that the Universe is now
expanding at an accelerated rate\cite{Riess07,Union2,PlanckXXVI}. It is usually assumed that this accelerated expansion is
caused by the so called dark energy (DE),  a cosmic fluid characterized by a negative  pressure,
$p_{de}$, and a negative equation of state $w<0$. According to the last estimates, more than $70\%$ of
matter-energy in the Universe is in the form of dark energy, so that the Universe is today in the dark energy dominant era.  Unfortunately the nature of dark energy is not known. A  huge amount of models of dark energy (DE) have been so far proposed. The first candidate for DE has been the cosmological constant, $\Lambda$, with a constant equation of state (EoS) parameter, $w=-1$. This model, although it generally fit quite well a large part of the current astronomical observations, shows some difficulties, foremost of theoretical nature: for example, it turns out that it is difficult  to reconcile the small present day value of DE density with the value predicted by the quantum field theories  \cite{Carroll}--\cite{Sahni}. From the observational point of view,  the present data seem to favor an evolving DE, with the EoS parameter allowed to cross the value $w=-1$ ({\it phantom crossing}  ). Moreover, a large part of the dark energy models so far proposed is based on the Einstein General Relativity (GR)  theory of gravity,  and it is therefore conceivable that  the  observed accelerating expansion could be interpreted as effects of alterations of the gravity action introduced in modified theories. Actually, in the last years different extensions to gravity have been widely investigated: by choosing  the gravitational action as function of the Ricci scalar, $f(R)$\cite{bla}--\cite{Capo3}, or considering the framework of $F(R,G)$ gravity in which  $F(R, G)$ is a generic function of the Ricci curvature scalar R and the Gauss-Bonnet topological invariant $G$ \cite{Maria}, or other curvature invariants\cite{bla2}. In a quite different approach, avoiding the curvature defined via the Levi-Civita connection, it is possible to introduce and use an alternative connection that has no curvature (vanishing $R$ ) but torsion $T$ ({\it teleparallelism gravity}). This has the property that the torsion is formed completely from products of first derivatives of the tetrad, with no second derivatives appearing in the torsion
tensor \footnote{The property of a vanishing R is motivated by the inflationary phase in the early stages of the universe evolution.}.
Thus, one can extend the Lagrangian density by adding a torsion function, i.e. $f(T )$. Our work, as the most part of  research in f(T) gravity, is naturally focused on the application  to cosmology, comparing the theoretical predictions to the observational data.It turns out that within these $f(T)$ models, it is possible to obtain an inflationary evolution  without an inflation and the {\it dark torsion} provides the accelerated expansion of the universe  \cite{Wu}--\cite{Yang2} . However, it worth noting that recently also astrophysical applications started to be discussed in literature (see for instance \cite{Bohmer}\, \cite{Wang11}\, and \cite{Abbas15} ) .
Just as in the case of  $f(R)$-gravity theories, it turns out that several models of the $f(T)$-gravity  agree with different data  \cite{Cardone}, so that these alternative approaches cannot  be discriminated so far from the {\it observational} point of view.  Moreover, a key topic in both the $f(R)$, and the $f(T)$ scenarios is the choice of  these functions. To overcome this degeneration, it is possible to adopt a quite different approach, the cosmography,  relying on  the time-series expansion of the scale factor. The cosmographic approach allows us to fit the data ( for instance on the distance - redshift relation), without any a priori assumption on the underlying cosmological model. It is just based on the assumption that the Universe  can be described by  the Friedmann-Lemaitre-Robertson-Walker metric.
The SNIa Hubble diagram extends up to $z = 1.7$ thus invoking the need for, at least, a fifth order Taylor expansion of the scale factor in order to give a reliable approximation of the distance - redshift relation. As a consequence, it could be, in principle, possible to estimate up to five cosmographic parameters, $(h, q_0, j_0, s_0, l_0)$, although the still too small data sets available does not allow to get a precise and realistic determination of all of them. Once these quantities have been determined, it is actually possible to relate the cosmographic parameters and the present day value of $f^{(n)}(T) = d^nf/dT^n$, with $n = 0,...,5$. These relations will allow to achieve reliable cosmological constraints on the $f(T)$ derivatives, and to test some  $f(T)$ gravity models.
For constraining the cosmographic parameters, we use the Union2 Type Ia Supernovae (SNIa) data set, the Hubble diagram constructed from some Gamma Ray Bursts luminosity distance indicators, and gaussian priors on the distance from the Baryon Acoustic Oscillations (BAO), and the Hubble constant $h$ (such priors have been included in order to help break the degeneracies among model parameters). 
 In Section II we will review the $f(T)$- gravity. In Section III we will introduce the basic notions of the cosmographic parameters. Section IV contains the main result of the paper demonstrating how the $f(T)$ derivatives can be related to the cosmographic parameters.We summarize and conclude in Section V.

\section{f(T) gravity: basics}
For the tangent space we consider an orthonormal basis $e_{A} (x^{\mu} ) $ ( vierbein)  at each point $x^{\mu}$ on a 4 dimensional manifold, where  $A=0, 1, 2, 3$,  in order to obtain the cosmological equations.
 Thus, the metric  is obtained from the dual basis $ e^A(x^{\mu})$:
\begin{equation}\label{base}
g_{\mu\nu}(x)=\eta_{ij} e^i_\mu(x)e^j_\nu(x)\,,
\end{equation}
being  $e^A_{\mu}$  the coordinates of the dual basis with respect to a coordinate basis on the cotangent space. Teleparallelism gravity uses  the so called Weitzenb\"{o}ck connection, to introduce a non-null torsion $T^\lambda_{\mu\nu}$ given by:
\begin{equation}\label{torsion}
    T^\lambda_{\mu\nu}=\hat{\Gamma}^\lambda_{\nu\mu}-\hat{\Gamma}^\lambda_{\mu\nu}=e^\lambda_i(\partial_\mu e^i_\nu - \partial_\nu e^i_\mu)\,.
\end{equation}
The
associated Lagrangian is
\begin{equation}\label{lagrangian}
    T={S_\rho}^{\mu\nu}{T^\rho}_{\mu\nu},
\end{equation}
being
\begin{equation}\label{s}
    {S_\rho}^{\mu\nu}=\frac{1}{2}({K^{\mu\nu}}_\rho+\delta^\mu_\rho {T^{\theta\nu}}_\theta-\delta^\nu_\rho {T^{\theta\mu}}_\theta)
\end{equation}
and ${K^{\mu\nu}}_\rho$ is the contorsion tensor
\begin{equation}\label{contorsion}
    {K^{\mu\nu}}_\rho=-\frac{1}{2}({T^{\mu\nu}}_\rho-{T^{\nu\mu}}_\rho-{T_\rho}^{\mu\nu}),
\end{equation}


\subsection{$f(T)$ cosmological equations}

{Our starting point is the {teleparallelism} action}
\begin{equation}\label{action}
   S = \frac{1}{16\pi G}\int{d^4x \sqrt{-g} f(T)}\,.
\end{equation}
The fields equations can are derived by variating the action  with respect to the vierbein leads: \cite{Ferraro}
\begin{eqnarray}
  && e^{-1}\partial_\mu(e{S_i}^{\mu\nu})f'(T)-e_i^\lambda {T^\rho}_{\mu\lambda}{S_\rho}^{\nu\mu}f'(T) +  \nonumber \\ & & +{S_i}^{\mu\nu}\partial_\mu(T)f''(T) +\frac{1}{4}e^\nu_if(T)=4\pi G{e_i}^\rho {T_\rho}^\nu, \label{equations}
\end{eqnarray}
where a prime denotes differentiation with respect to $T$,
${S_i}^{\mu\nu}={e_i}^\rho {S_\rho}^{\mu\nu}$ and $T_{\mu\nu}$ is the energy-momentum
tensor.

{ Under the assumption of a flat FRW universe we have that}
\begin{equation}\label{metric}
    e^i_\mu = diag(1, a(t), a(t), a(t)),
\end{equation}
where $a(t)$ is the cosmological scale factor. By using Eqs. (\ref{torsion}), (\ref{lagrangian}), (\ref{s}) and (\ref{contorsion}) we obtain that
\begin{equation}\label{lt}
    T=-6H^2,
\end{equation}
where $H=\frac{\dot a}{a}$ is the Hubble function. By substituting  the vierbein (\ref{metric}) in (\ref{equations})
for $i=0=\nu$ we obtain the $f(T)$ Friedmann equations:
\begin{equation}\label{friedmann}
    12H^2f'(T)+f(T)=16\pi G\rho.
\end{equation}
The equation $i=1=\nu$ is
\begin{equation}\label{acceleration}
    48H^2f''(T)\dot{H}-f'(T)[12H^2+4\dot{H}]-f(T)=16\pi Gp.
\end{equation}
Here $\rho$ and $p$ are the dark matter energy density and pressure, respectively, which  satisfy the conservation equation
\begin{equation}\label{conservation}
    \dot{\rho}+3H(\rho+p)=0.
\end{equation}
{ We can formally rewrite Eqs. (\ref{friedmann}) and (\ref{acceleration})  in the standard form, introducing an effective torsion contributions to the energy density and pressure $  \rho_T$ and $ \rho_T$:}
\begin{equation}\label{modfri}
    H^2=\frac{8\pi G}{3}(\rho+\rho_T),
\end{equation}
\begin{equation}\label{modacce}
    2\dot H+3H^2=-\frac{8\pi G}{3}(p+p_T)
\end{equation}
where
\begin{equation}\label{rhoT}
    \rho_T=\frac{1}{16\pi G}[2Tf'(T)-f(T)-T/2],
\end{equation}
\begin{equation}\label{pT}
    p_T=\frac{1}{16\pi G}[2\dot H(4Tf''(T)+2f'(T)-1)]-\rho_T.
\end{equation}
 Therefore, by using Eqs. (\ref{rhoT}) and (\ref{pT}), we can define
the effective torsion equation of state
\begin{equation}\label{omegaeff}
    \omega_{T}\equiv\frac{p_T}{\rho_T}=-1+\frac{4\dot H(4Tf''(T)+2f'(T)-1)}{4Tf'(T)-2f(T)-T}\,,
\end{equation}
which can be related to the observed accelerated rate of the Universe expansion.

\section{Cosmography}

The cosmographic approach to cosmology recently aroused great interest as it allows to obtain information from observations (mainly measured distances)  just assuming  the minimal priors of isotropy and homogeneity, without assuming which kind of dark energy and dark matter are needed to satisfy the Einstein equation. 
From isotropy and homogeneity priors the space-time geometry is described by the  FLRW line element
\begin{equation}
ds^2=c^2dt^2-a^2(t)\left[\frac{dr^2}{1-kr^2}+r^2d\Omega^2\right]\,,
\end{equation}
where $a(t)$ is the scale factor and $k= +1, 0, - 1$ is the curvature parameter.
Using this metric, it is possible to express the luminosity
distance $d_L$ as a power series in the redshift parameter $z$ whose coefficients are functions of the scale factor
$a(t)$ and its higher order derivatives. This
expansion leads to a distance\,-\,redshift relation that is fully model independent since it does not depend on the
particular form of the solution of cosmic evolution equations. To this purpose,
it is convenient to introduce the \textbf{cosmographic functions} \cite{Visser}:
\begin{eqnarray}
\label{eq:cosmopar}
H(t) &\equiv& + \frac{1}{a}\frac{da}{dt}\, ,
\\
q(t) &\equiv& - \frac{1}{a}\frac{d^{2}a}{dt^{2}}\frac{1}{H^{2}}\,
,
\\
j(t) &\equiv& + \frac{1}{a}\frac{d^{3}a}{dt^{3}}\frac{1}{H^{3}}\,
,
\\
s(t) &\equiv& + \frac{1}{a}\frac{d^{4}a}{dt^{4}}\frac{1}{H^{4}}\,
,
\\
l(t) &\equiv& + \frac{1}{a}\frac{d^{5}a}{dt^{5}}\frac{1}{H^{5}}\,.
\end{eqnarray}
The cosmographic parameters, which are usually indicated as the \textit{Hubble},
\textit{deceleration}, \textit{jerk}, \textit{snap} and
\textit{lerk} parameters, respectively, correspond to the functions evaluated at the present time $t_0$ \footnote{Note that the
use of the jerk parameter to discriminate between different models was also
proposed in \cite{SF} in the context of the {\it statefinder}
parametrization.}. Furthermore, it is possible
to relate the derivative of the Hubble parameter to the
cosmographic parameters\,:
\begin{eqnarray}
\dot{H} &=& -H^2 (1 + q) \ , \label{eq: hdot}
\\
\ddot{H} &=& H^3 (j + 3q + 2) \ , \label{eq: h2dot}
\\
d^3H/dt^3 &=& H^4 \left ( s - 4j - 3q (q + 4) - 6 \right ) \ ,
\label{eq: h3dot}
\\
d^4H/dt^4 &=& H^5 \left ( l - 5s + 10 (q + 2) j + 30 (q + 2) q +
24 \right ) \ , \label{eq: h4dot}
\end{eqnarray}
where a dot denotes derivative with respect to the cosmic time
$t$.
It turns out  that equivalent expressions can be obtained  starting from the Taylor
series expansion of the Hubble parameter{\setlength\arraycolsep{0.2pt}}
\begin{eqnarray}\label{eq:Hseriesdef}
H(z) &=& H_{0} + \frac{dH}{dz}\Bigg{|}_{z=0} z + \frac{1}{2!}
\frac{d^{2}H}{dz^{2}}\Bigg{|}_{z=0} z^{2} + \frac{1}{3!}
\frac{d^{3}H}{dz^{3}}\Bigg{|}_{z=0} z^{3} +
\frac{1}{4!} \frac{d^{4}H}{dz^{4}}\Bigg{|}_{z=0} z^{4} +
\emph{O}(z^{5}) \, ,
\end{eqnarray}
and using the derivation rule
\begin{equation}\label{eq:firstHt}
\frac{d}{dt} = -(1 + z) H \frac{d}{dz}\,.
\end{equation}


\subsection{f(T) cosmography}
In order to relate the present day values of $f(T)$ and its derivatives to the cosmographic parameters $(q_0, j_0, s_0, l_0)$ the first step consists in to differentiating Eq. (\ref{lt}).with respect to $t$, so that:
\begin{eqnarray}
  \dot T &=& -12H\dot H, \label{dif1T} \\
  \ddot T &=& -12[\dot H^2+H\ddot H], \\ \label{dif2T}
  \dddot T &=& -12[3\dot H\ddot H+H\dddot H], \\ \label{dif3T}
  T^{(iv)} &=& -12[3\ddot H^2+4\dot H\dddot H+HH^{(iv)}] \label{dif4T}
\end{eqnarray}

Therefore we can rewrite the Friedmann Eqs. (\ref{friedmann}) and (\ref{acceleration}):
\begin{equation}\label{friedfried}
    H^2=\frac{-1}{12f'(T)}[T\Omega_{m}+f(T)]
\end{equation}
and
\begin{equation}\label{acceacce}
    \dot H=\frac{1}{4f'(T)}[T\Omega_{m}-4H\dot{T}f''(T)]
\end{equation}
 However, in order to enter other cosmographic parameters we have to further differentiate  Eq.(\ref{acceacce}):


\begin{equation}\label{expddH}
    \ddot H=\frac{\Omega_m}{4Hf'(T)}[H\dot T-T(3H^2+2\dot H)]-\frac{1}{f'(T)}[(2\dot H\dot T+H\ddot T)f''(T)+H\dot T^2f'''(T)],
\end{equation}
\begin{eqnarray}
   \dddot H&=&\frac{\Omega_m}{4H^2f'(T)}[T(9H^4+6H^2\dot H+4\dot H^2)-H\dot T(3\dot H+6H^2)+H(H\ddot T-2\ddot H T)] \nonumber \\ &-&\frac{1}{Hf'(T)}[\dot H\ddot Hf'(T)+(2\dot H^2\dot T+3H\ddot H\dot T+4H\dot H\ddot T+H^2\dddot T)f''(T)+H^2\dot T^3f^{(iv)}(T) \nonumber \\&+&H\dot T(4\dot H\dot T+3H\ddot T)f'''(T)], \label{expdddH}
\end{eqnarray}


The above Eqs.(36) (37), evaluated at the present day, can be used to constrain  the  unknown quantities $f(T_0)$, $f'(T_0)$, $f''(T_0)$, $f'''(T_0)$,and  $f^{(iv)}(T_0)$.
 in $f(T)$ gravity we have  an effective (time dependent) gravitational constant $G_{eff}=\frac{G_N}{f'(T)}$. Assuming that the present day value of $G_{eff}$ is the same as the Newtonian gravitational constant, we have the simple constraint:
\begin{equation}\label{constraint}
  f'(T_0)=1 \cdot
\end{equation}
Assuming  that $f(T)$ a Taylor expansion of $f(T)$ in $T-T_0$, after some algebra it turns out that(see also\cite{vincft})

\begin{equation}
f_0= \frac{f(T_0)}{6H_0^2}=\Omega_{m0}-2,
\label{f0T}
\end{equation}
\begin{equation}\label{f1T}
 f_1=   f'(T_0)=1, 
\end{equation}
\begin{equation}\label{f2T}
  f_2=  \frac{f''(T_0)}{(6H_0^2)^{-1}}=\frac{-3\Omega_{m0}}{4(1+q_0)}+\frac{1}{2},
 \end{equation}
\begin{equation}\label{f3T}
f_3=    \frac{f'''(T_0)}{(6H_0^2)^{-2}}=\frac{-3\Omega_{m0}(3q_0^2+6q_0+j_0+2)}{8(1+q_0)^3}+\frac{3}{4},
\end{equation}
\begin{eqnarray}\label{f4T}
  f_4=  \frac{f^{(iv)}(T_0)}{(6H_0^2)^{-3}}&=&\frac{-3\Omega_{m0}}{16(1+q_0)^5}[s_0(1+q_0)+j_0(6q_0^2+17q_0+3j_0+5) \nonumber \\&+&3q_0(5q_0^3+20q_0^2+29q_0+16)+9]+\frac{15}{8},
\end{eqnarray}

\subsection{Statistical analysis}
To fit the present cosmographic parameters $(q_0, j_0, s_0, l_0)$ using the observational data we need the cosmographic expansion of some observational quantities, which are the luminosity distance and the angular-diameter distance, evaluable as

\begin{equation}
 d_{L} = (1+z)\: ( r(z))\,,
 \end{equation}
 \begin{equation}
 d_{A} = \frac{1}{1+z} \: ( r(z))\,,
  \end{equation}
where $r_{0}$ is:
\begin{equation}\label{eq:r_sin}
r(z) = \left\{
\begin{array}{lr}
  \sin ( \int_{0}^{z} \frac{1 }{H(z^{'})} dz^{'} ) &  k = +1; \\
  &  \\
 \int_{0}^{z} \frac{1 }{H(z^{'})} dz^{'}&  k = 0; \\
  &  \\
  \sinh ( \int_{0}^{z} \frac{1 }{H(z^{'})} dz^{'}) &  k = -1.
\end{array} \right.
\end{equation}
{ Actually the cosmic expansion history of the  Universe is a fundamental element for  understanding  the physics driving the cosmological dynamics, and determining the nature of  the Universe underlying its own  dark components, mainly the nature of the dark energy. Already distance measurements have shown that the cosmological constant could explain the observed acceleration.  What remains to be proved is whether the dark energy is constant,  or varies in redshift and perhaps space coordinates, or whether describing the Universe by bare General Relativity fails on large scale,  and  some extended theory of gravity is needed. Whatever future observations will discover, the  basic plot of the distance scale will be a primary icon of the modern cosmology. The data for this plot come primarily from measurements of SNeIa and baryon acoustic oscillations, which show a deep complementarity: actually SNeIa measurements have negligible cosmic variance limit; that is, under a practical approach, in theory we can build  up arbitrarily large samples. This allows us to measure relative distances at redshift $z\leq 1.5$, where the cosmic volume limits BAO measurements, which offer comparable precision above $1<z$ and provide an absolute calibration that ties the measurement of the CMB (Cosmic Microwave Background)shift parameter at last scatter redshift $z_{*}=1090.10$. A combined analysis of CMB, BAO and SNeIa allows us to calibrate the distance scale from $z=0$ to $z\simeq 1000$.  The GRB Hubble diagram  cover a  range redshift range $ 0.1\leq z \leq 9 $ which is not spanned by the SNeIa data, neither by BAOs, and which is very important to  investigate the dark energy  equation of state, and test different cosmological models.}

It turns out that the cosmographic expansion of the luminosity distance is given as:
\begin{eqnarray}\label{serielum1}
d_{L}(z) = \frac{c z}{H_{0}} \left( \mathcal{D}_{L}^{0} +
\mathcal{D}_{L}^{1} \ z + \mathcal{D}_{L}^{2} \ z^{2} +
\mathcal{D}_{L}^{3} \ z^{3} + \mathcal{D}_{L}^{4} \ z^{4} +
\emph{O}(z^{5}) \right)\,,
\end{eqnarray}
with: {\setlength\arraycolsep{0.2pt}
\begin{eqnarray}\label{serieslum2}
\mathcal{D}_{L}^{0} &=& 1\,, \\
\mathcal{D}_{L}^{1} &=& - \frac{1}{2} \left(-1 + q_{0}\right)\,, \\
\mathcal{D}_{L}^{2} &=& - \frac{1}{6} \left(1 - q_{0} - 3q_{0}^{2} + j_{0}\right)\,, \\
\mathcal{D}_{L}^{3} &=& \frac{1}{24} \left(2 - 2 q_{0} - 15
q_{0}^{2} - 15 q_{0}^{3} + 5 j_{0} + 10 q_{0} j_{0} + s_{0} \right)\,,\\
\mathcal{D}_{L}^{4} &=& \frac{1}{120} \left( -6 + 6 q_{0} + 81
q_{0}^{2} + 165 q_{0}^{3} + 105 q_{0}^{4} - 110 q_{0} j_{0} - 105
q_{0}^{2} j_{0} - 15 q_{0} s_{0} + \right.\nonumber \\
&-& \left.  27 j_{0} + 10 j^{2} - 11 s_{0} - l_{0}\right)\,.
\end{eqnarray}}
For the angular diameter distance we have:
\begin{eqnarray}
d_{A}(z) = \frac{c z}{H_{0}} \left ( \mathcal{D}_{A}^{0} +
\mathcal{D}_{A}^{1}  z + \mathcal{D}_{A}^{2} \ z^{2} +
\mathcal{D}_{A}^{3}  z^{3} + \mathcal{D}_{A}^{4}  z^{4} +
\emph{O}(z^{5}) \right )\,,
\end{eqnarray}
with: {\setlength\arraycolsep{0.2pt}
\begin{eqnarray}
\mathcal{D}_{A}^{0} &=& 1 \,,\\
\mathcal{D}_{A}^{1} &=& - \frac{1}{2} \left(3 + q_{0}\right)\,, \\
\mathcal{D}_{A}^{2} &=& \frac{1}{6} \left(11 + 7 q_{0} + 3q_{0}^{2} - j_{0} \right)\,, \\
\mathcal{D}_{A}^{3} &=& - \frac{1}{24} \left(50 + 46 q_{0} + 39
q_{0}^{2} + 15 q_{0}^{3} - 13 j_{0} - 10 q_{0} j_{0} - s_{0}
- \frac{2 k c^{2} (5 + 3 q_{0})}{H_{0}^{2} a_{0}^{2}}\right)\,, \\
\mathcal{D}_{A}^{4} &=& \frac{1}{120} \left( 274 + 326 q_{0} + 411
q_{0}^{2} + 315 q_{0}^{3} + 105 q_{0}^{4} - 210 q_{0} j_{0} - 105
q_{0}^{2} j_{0} - 15 q_{0} s_{0} + \right. \nonumber \\
&-& \left. 137 j_{0} + 10 j^{2} - 21 s_{0} - l_{0} \right)\,.
\end{eqnarray}}
To fit the cosmographic parameters and derive their own confidence regions, we use the currently updated sample of SNIa and GRB Hubble Diagrams,  and we assume  gaussian priors on the distances from the Baryon Acoustic Oscillations (BAO), and the Hubble constant $h$. Such priors have been included in order to help break the degeneracies among the parameters of the cosmographic series expansion in Eqs. (\ref{serielum1}).
\subsubsection{Supernovae}
\label{sec:SNdata}
 Over the last decade the confidence in type Ia supernovae as standard candles has been steadily
growing;
the luminosity at peak brightness of a supernova can be inferred from the shape and wavelenght - dependence of its flux evolution. From this luminosity is possible to derive the luminosity distance which is related to the cosmological parameters, including the dark energy properties.
 Here we consider the recently updated Supernovae Cosmology Project \textit{Union 2.1} compilation \cite{Union2}, which is an update of the original \textit{Union2} compilation, consisting of $580$ SNIa, spanning the redshift range ($0.015 \le z \le 1.4$).
We actually
compare the \textit{theoretically\thinspace\ predicted} distance modulus $\mu(z)$
with the \textit{observed} one, through a Bayesian approach, based on the definition
of the cosmographic distance modulus,

\begin{equation}
\mu(z_{j}) = 5 \log_{10} ( D_{L}(z_{j}, \{\theta_{i}\}) )+\mu_0\,,
\end{equation}
where $D_{L}(z_{j}, \{\theta_{i}\})$ is the Hubble free luminosity
distance, expressed as a series depending on the cosmographic parameters, $\theta_{i}=(q_{0},
j_{0}, s_{0}, l_{0})$.  The parameter $\mu_{0}$ encodes the Hubble
constant and the absolute magnitude $M$, and has to be
marginalized over. Giving the heterogeneous origin of the Union data
set, we have worked with an alternative version of  the $\chi^2$:
\begin{equation}\label{eq: sn_chi_mod}
\tilde{\chi}^{2}_{\mathrm{SN}}(\{\theta_{i}\}) = c_{1} -
\frac{c^{2}_{2}}{c_{3}}\,,
\end{equation}
being
\begin{equation}
c_{1} = \sum^{{\cal{N}}_{SNIa}}_{j = 1} \frac{(\mu(z_{j}; \mu_{0}=0,
\{\theta_{i})\} -
\mu_{obs}(z_{j}))^{2}}{\sigma^{2}_{\mathrm{\mu},j}}\, ,
\end{equation}
\begin{equation}
c_{2} = \sum^{{\cal{N}}_{SNIa}}_{j = 1} \frac{(\mu(z_{j}; \mu_{0}=0,
\{\theta_{i})\} -
\mu_{obs}(z_{j}))}{\sigma^{2}_{\mathrm{\mu},j}}\, ,
\end{equation}
\begin{equation}
c_{3} = \sum^{{\cal{N}}_{SNIa}}_{j = 1}
\frac{1}{\sigma^{2}_{\mathrm{\mu},j}}\,.
\end{equation}
It is worth noting that
\begin{equation}
\chi^{2}_{\mathrm{SN}}(\mu_{0}, \{\theta_{i}\}) = c_{1} - 2 c_{2}
\mu_{0} + c_{3} \mu^{2}_{0} \,.
\end{equation}
\subsubsection{The Gamma Ray Bursts Hubble diagram}
GRBs are visible up to high $ z $, thanks to the enormous energy that they release, and thus may be good candidates  for our high-redshift cosmological investigation. Sadly, GRBs may be everything but standard candles since their peak luminosity spans a wide range, even if  there have been many   efforts to make them standardizable candles using
some empirical correlations among distance dependent quantities and rest frame observables  \cite{Amati08}.
 These empirical relations allow one to deduce the GRB rest frame luminosity or energy from an observer
frame measured quantity so that the distance modulus can be obtained with an error which depends essentially on the intrinsic scatter of the adopted correlation.
In this paper we use a GRBs Hubble Diagram data set, build up by calibrating the $E_{\rm p,i}$ -- $E_{\rm iso}$ relation
Union SNIa sample,  \cite{ME,MELMS} . Here we consider only $98$ GRBs with $1.\leq z \le 4$  in order to cover the a redshift range which is not spanned by the SNeIa data.
\subsubsection{Baryon Acoustic Oscillations}
Baryon Acoustic Oscillations data are promising standard rulers to investigate different cosmological scenarios and models. They  are related to density fluctuations induced by acoustic waves created by primordial perturbations:  the peaks of the acoustic waves gave rise to dense regions of baryons which, at recombination,  imprint the correlation between matter densities at the scale of the sound horizon.  Measurements of CMBR provide the absolute physical scale for these baryonic peaks, but the observed position of the peaks of the two-point correlation function of the matter distribution, compared with such absolute values, makes possible  measurements of cosmological distance scales. 
{ Using the BAO as standard rules it's possible to measure the angular diameter distance and the Hubble parameter as functions of redshift.}
In order to use BAOs as cosmological tool, we follow \cite{P10}, defining\,:
\begin{equation}
d_z = \frac{r_s(z_d)}{d_V(z)}\,,
\label{eq: defdz}
\end{equation}
where $z_d$ is the drag redshift computed according to the formula in \cite{EH98}, $r_s(z)$ is the comoving sound horizon\,:
\begin{equation}
r_s(z) = \frac{c}{\sqrt{3}} \int_{0}^{(1 + z)^{-1}}{\frac{da}{a^2 H(a) \sqrt{1 + (3/4) \Omega_b/\Omega_{\gamma}}}} \ ,
\label{defsoundhor}
\end{equation}
and $d_V(z)$ the volume distance, given by\,:

 \begin{equation}
 d_V(z,{\rm\theta}) = \left[\left(1+z\right) d_A(z,{\rm\theta})^2\frac{c z}{H(z,{\rm\theta})}\right]^{\frac{1}{3}}.\label{volumed}
\end{equation}
Here  $ d_L(z,\mathbf{\theta})$ is the luminosity distance and $ d_A(z,{\rm\theta})$ the angular diameter distance.

\subsection{Likelihood analysis}
In order to constrain the cosmographic parameters, we perform a preliminary procedure to maximize the likelihood function ${\cal{L}}({\bf p}) \propto \exp{[-\chi^2({\bf p})/2]}$, where ${\bf p}$ is the set of cosmographic parameters and the expression for $\chi^2({\bf p})$ depends on the data set used.  For this test we consider only the SNIa data, thus we define\,:
\begin{eqnarray}
\chi^2({\bf p}) & = & \sum_{i = 1}^{{\cal{N}}_{SNIa}}{\left [ \frac{\mu_{obs}(z_i) - \mu_{th}(z_i, {\bf p})}{\sigma_i} \right ]^2} \nonumber \\
& + &  \left ( \frac{h - 0.742}{0.036} \right )^2
+ \left ( \frac{\omega_m - 0.1356}{0.0034} \right )^2 \,,
\label{defchiSNIa}
\end{eqnarray}
being, $\mu_{obs}$ and $\mu_{th}$ are the observed and theoretically predicted values of the distance modulus, while the sum is over all the SNIa  in the sample. The last two terms are Gaussian priors on $h$ and $\omega_M = \Omega_M h^2$ and are included in order to help break the degeneracies among the cosmographic parameters. To this aim, we have resorted to the results of the SHOES collaboration \cite{shoes} and the WMAP7 data \cite{WMAP7}, respectively, to set the numbers used in Eqs. (\ref{defchiSNIa}).
When we are using GRBs only, we define\,:
\begin{eqnarray}
\chi^2({\bf p}) & = & \sum_{i = 1}^{{\cal{N}}_{GRBHD}}{\left [ \frac{\mu_{obs}(z_i) - \mu_{th}(z_i, {\bf p})}{\sigma_i} \right ]^2} \nonumber \\
& + &  \left ( \frac{h - 0.742}{0.036} \right )^2
+ \left ( \frac{\omega_m - 0.1356}{0.0034} \right )^2 \ .
\label{chigrb}
\end{eqnarray}

We finally combine the SNIa and GRBs datasets  with other data redefining ${\cal{L}}({\bf p})$ as\,:
\begin{eqnarray}
{\cal{L}}({\bf p}) & \propto & \frac{\exp{(-\chi^2_{SNIa/GRB}/2)}}{(2 \pi)^{\frac{{\cal{N}}_{SNIa/GRB}}{2}} |{\bf C}_{SNIa/GRB}|^{1/2}} \nonumber \\
~ & \times  & \frac{1}{\sqrt{2 \pi \sigma_h^2}} \exp{\left [ - \frac{1}{2} \left ( \frac{h - h_{obs}}{\sigma_h} \right )^2
\right ]} \nonumber \\
~ & \times & \frac{\exp{(-\chi^2_{BAO}/2})}{(2 \pi)^{{\cal{N}}_{BAO}/2} |{\bf C}_{BAO}|^{1/2}} \nonumber \\
~ & \times & \frac{1}{\sqrt{2 \pi \sigma_{{\cal{R}}}^2}} \exp{\left [ - \frac{1}{2} \left ( \frac{{\cal{R}} - {\cal{R}}_{obs}}{\sigma_{{\cal{R}}}} \right )^2 \right ]} \nonumber \\
~ & \times & \frac{\exp{(-\chi^2_{H}/2})}{(2 \pi)^{{\cal{N}}_{H}/2} |{\bf C}_{H}|^{1/2}} \  .
\label{defchiall}
\end{eqnarray}

The first two terms are the same as defined above, where ${\bf C}_{SNIa/GRB}$ the SNIa/GRBs diagonal covariance matrix and $(h_{obs}, \sigma_h) = (0.742, 0.036)$. The third term takes into account the BAO constraints on $d_z = r_s(z_d)/D_V(z)$. We set $r_s(z_d) = 152.6 \ {\rm Mpc}$ from PLANCK, and use the SDSS galaxy sample data.
The values of $d_z$ at $z = 0.20$ and $z = 0.35$ have been indeed estimated by Percival et al. (Ref.~\refcite{P10}), so that we define $\chi^2_{BAO} = {\bf D}^T {\bf C}_{BAO}^{-1} {\bf C}$ with ${\bf D}^T = (d_{0.2}^{obs} - d_{0.2}^{th}, d_{0.35}^{obs} - d_{0.35}^{th})$ and ${\bf C}_{BAO}$ is the BAO covariance matrix. The term $\displaystyle  \frac{1}{\sqrt{2 \pi \sigma_{{\cal{R}}}^2}} \exp{\left [ - \frac{1}{2} \left ( \frac{{\cal{R}} - {\cal{R}}_{obs}}{\sigma_{{\cal{R}}}} \right )^2 \right ]} $ in Eq. (\ref{defchiall}) considers the shift parameter${\cal{ R}}$:

\begin{equation}
{\cal{R}} = H_{0} \sqrt{\Omega_M} \int_{0}^{z_{\star}}{\frac{dz'}{H(z')}}\,,
\label{eq: defshiftpar}
\end{equation}
where $z_\star = 1090.10$ is the redshift of the last scattering surface  \cite{B97,EB99}\,. According to the WMAP7 data setting $({\cal{R}}_{obs}, \sigma_{{\cal{R}}}) = (1.725, 0.019)$.

\begin{figure}
\includegraphics[width=8 cm, height=6cm]{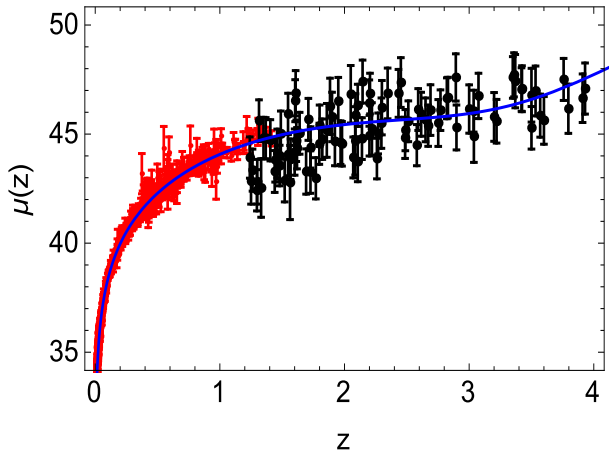}
\caption{The {\it extended } Hubble diagram obtained by joining the SNIa (red points) and the GRBs (black points) datasets and, superimposed, the theoretical models of distance corresponding to the best fit values of the parameters (blue solid line). } \label{HD}
\end{figure}

\begin{figure}
\includegraphics[width=8 cm, height=6cm]{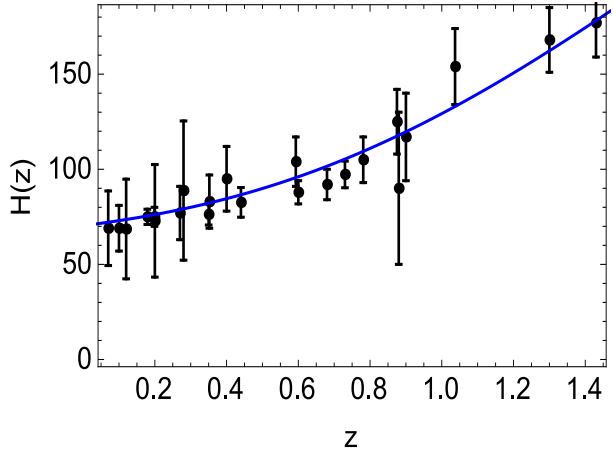}
\caption{The measured values of $H(z)$ and, superimposed, the theoretical function corresponding to the best fit values of the parameters (blue solid line).
} \label{hz}
\end{figure}

Finally, the term  $\displaystyle \frac{\exp{(-\chi^2_{H}/2})}{(2 \pi)^{{\cal{N}}_{H}/2} |{\bf C}_{H}|^{1/2}}  $ in Eq. (\ref{defchiall}) takes  some recent measurements of $H(z)$ from the differential age of passively evolving elliptical galaxies into account. We, actually, use the data collected by Farooq and Ratra (2013) giving the values of the Hubble parameter for ${\cal{N}}_H = 28$ different points over the redshift range $0.07 \le z \le 2.3$ with a diagonal covariance matrix.
In order to efficiently sample the ${\cal{N}} $ dimensional space of the cosmographic parameters, we use the  MCMC method running five parallel chains and using the Gelmann - Rubin diagnostic approach to test the convergence. It  uses parallel chains with dispersed initial values to check whether they all converge to the same target distribution.
As a test instrument it uses the reduction factor \textit{R}, which is the square root of the ratio of the variance between-chain  and the variance within-chain.  A large \textit{R} indicates that the between-chain variance is substantially greater than the within-chain variance, so that longer simulation is needed.
We want that  \textit{R}  converges to 1 for each parameter. We set $R - 1$ of order $0.05$, which is more restrictive than the often used and recommended value $R - 1 < 0.1$ for standard cosmological investigations.
Moreover in order to reduce the uncertainties on the cosmographic  parameters, since methods like the MCMC are based on an algorithm that moves randomly in the parameter space, we \textit{a priori} impose some basic consistency constraints on the positiveness of $H^{2}(z)$ and $d_{L}(z)$.
We eliminate first $30\%$  of the points iterations at the beginning of any MCMC run, when the Chain is far from the convergence,  and we thus join the many times -run chains. The histograms of the parameters from the merged chain are then used to infer median values and confidence ranges. Actually, the  confidence levels are estimated
from the final sample (the merged chain): the $15.87$-th and $84.13$-th quantiles define the $68 \%$ confidence interval; the $2.28$-th and $97.72$-th quantiles define the $95\%$ confidence interval;  and the $0.13$-th and $99.87$-th quantiles
define the $99\%$ confidence interval. In Table \ref{tab1} we present the results of our analysis.We consider that only the deceleration parameter, $q_0$, and the jerk, $j_0$, are well constrained:  indeed it turns out that $s_0$ is weakly constrained, and $l_0$ is unconstrained. In Figs. (\ref{HD}), (\ref{hz}) we show the best if curves with the corresponding datasets.

\begin{table}
\tbl{Constraints on the cosmographic  parameters combining the SNIa and GRBs  Hubble Diagrams (HD) with BAO and $H(z)$  data set.}
{\begin{tabular}{@{}cccccc@{}} \toprule
  Parameter &$h$&$q_0$&$j_0$&$s_0$&$l_0$\\
  \colrule
  Best Fit & $0.69$&$-0.34$ & $-0.65$ & $-3.3$ & $18.5$\\
  Mean & $0.68$&$ -0.32$&$-0.54$&$-2.8$&$ 17.5$\\
  2 $\sigma$ & $ (0.67, 0.71)$&$ (-0.6, -0.16)$&$(-1.03,0.90)$&$(-5.6, 1.5)$&$(8.8, 23.9)$\\
 \botrule
\end{tabular} \label{tab1}}
\end{table}
It is worth noting that,  because of the contribution of the GRBs HD and the $H(z)$ data, our statistical analysis  provides a value of the Hubble constant $h$ which is fully consistent with the Planck results.

\section{Implications for the f(T) gravity}
\begin{figure}
\includegraphics[width=6 cm, height=6cm]{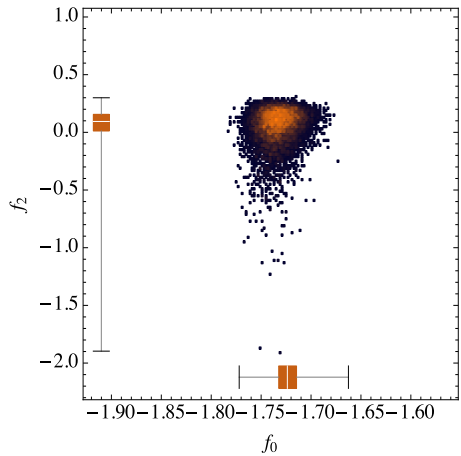}
\caption{Confidence regions in the($f_0$--$f_2$)  plane, as provided by  our analysis.  On the axes are plotted also the box-and-whisker diagrams relatively to the  respective parameters: the bottom and top of the diagrams are the 25th and 75th percentile (the lower and upper quartiles, respectively), and the band near the middle of the box is the median.} \label{f0f2}
\end{figure}

\begin{figure}
\includegraphics[width=6 cm, height=6cm]{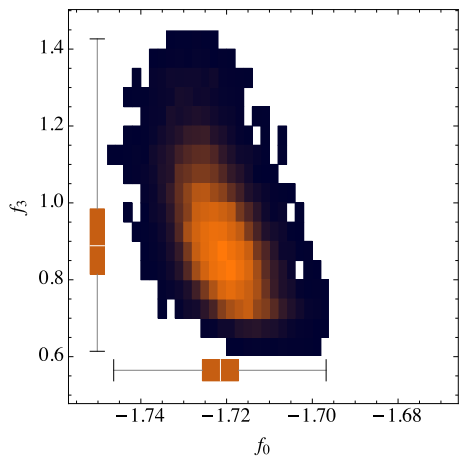}
\caption{Confidence regions in the($f_0$--$f_3$)  plane, as provided by  our analysis.  On the axes are plotted also the box-and-whisker diagrams relatively to the  respective parameters: the bottom and top of the diagrams are the 25th and 75th percentile (the lower and upper quartiles, respectively), and the band near the middle of the box is the median. The fuzzyness depends on numerical artifacts. } \label{f0f3}
\end{figure}

In this section  we want to relate the constraints on the cosmographic parameters to the derivatives of $ f(T )$ in order to set constraints on the model without any a priori assumption on the function $f (T )$. In order to perform this, we have to use the Eqs. (\ref{f0T}, ...,\ref{f4T}), which map the space of parameters $\left\{h,q_0,j_0,s_0\right\}$, in the space $\left\{f(T_0)\,,f'(T_0),f^{''}(T_0),f^{'''}(T_0),f^{(iv)}(T_0)\right\}$. The results of our analysis are shown in Table \ref{tab2} ; in Figs (\ref{f0f2}), and (\ref{f0f3}) is indeed shown the joint probability for different couples of $\left\{f(T_0)\right\}$, and its derivatives.  In the following we define $ f(T_0)=f_0$; and $\frac{f^{i}(T_0)}{{6 H_0^2}^{-(i-1)}}=f_{i}$

\begin{table}
\tbl{Constraints on the cosmographic  parameters combining the SNIa and GRBs  HDs with BAO and $H(z)$  data set.}
{\begin{tabular}{@{}ccccc@{}} \toprule
Function &$\frac{f(T_0)}{6 H_0^2}$&$\frac{f^{''}(T_0)}{{\left(6 H_0^2\right)}^{-1}}$&$\frac{f^{'''}(T_0)}{{\left(6 H_0^2\right)}^{-2}}$&$\frac{f^{iv}(T_0)}{{\left(6 H_0^2\right)}^{-3}}$\\
  \colrule
    Best Fit & $-1.71$&$0.18$ & $0.78$ & $2.13$ \\
  Mean & $-1.72$&$ 0.19$&$0.79$&$1.98$\\
  2 $\sigma$ & $ (-1.73,-1.67)$&$ (-0.02, 0.26)$&$(0.13,1.17)$&$(1.25, 5.2)$\\
 \botrule
\end{tabular} \label{tab2}}
\end{table}
We limit our cosmographic constraints to  $f^{(iv)}(T_0)$, since it turns out that $f^{(v)}(T_0)$ would be actually unconstrained, depending on both $s_0$ and $l_0$. Moreover,
it is worth noting that the best fit value  $\left\{f(T_0)\right\}=-1.72$ provides a value for the dark matter density parameter $\Omega_m= 0.28$, in agreement with recent independent estimation (see for instance  \cite{PlanckXXVI}).

\subsection{Connection with  $f(T)$ cosmological models}

In this section we illustrate the possibility to test the reliability of a certain $f(T)$ cosmological model without solving the field equations and using the observational data to set  its characterizing parameters: indeed the can chosen by requiring that  the cosmographic constraints  are satisfied. In order to illustrate the method we consider some of the most common $f(T)$ cosmological models:

\begin{romanlist}[(b)]
\item  the power-law functional  first discussed by Bengochea et al. and Wu et al.(See Ref.~\refcite{Bengo},Ref.~\refcite{Wu3}):
\begin{eqnarray}
&&f(T) = \alpha ( - T)^b  \,;\label{ftBengochea}\\
\end{eqnarray}

\item power law and {\it modified} power law models  discussed by Myrzakulov et al.(See Ref.~\refcite{Myrza2}):
\begin{eqnarray}
&&f(T) = \alpha T + \beta T^n\,,\label{ftmyrzabis}\\
&&f(T) = \alpha T + \beta T^{\delta} \ln{T}\,. \label{ftmyrza}
\end{eqnarray}

\item A $\tanh $ f(T) model models discussed by Juan et al. (See Ref.~\refcite{Hong2}):
\begin{eqnarray}
&&f(T)=\alpha  T^n \tanh \left(\frac{T_{0}}{T}\right)\,,\label{wuhongewi1}\\
\end{eqnarray}
\item a  $\log$ f(T) model discussed by  Bamba et al.(See Ref.~\refcite{Bamba}):
\begin{eqnarray}
f(T)=\alpha  T_{0} \sqrt{\frac{\beta\,  T_{0}}{T}} \log \left(\frac{\beta T_{0}}{T}\right)\,.
\label{Bamba}
\end{eqnarray}
\end{romanlist}

\begin{figure}
\includegraphics[width=8 cm, height=7cm]{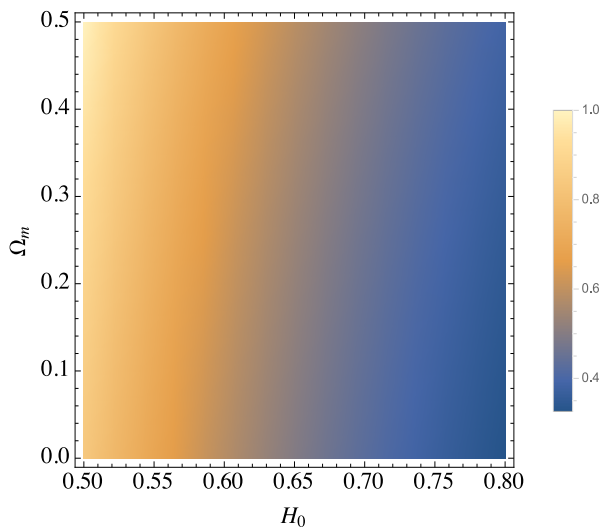}
\caption{Density plot in the($h$--$\Omega_m$)  plane for the derivative $f_2$ in the case of the power-law functional  (\ref{ftBengochea}) . It turns out that for $\Omega_ m$  varying in the range $\left(0.2,0.33\right)$ and $h$ in the range $\left(0.65, 0.7\right)$  $f_2$  and $f_3$  are in agreement with the  cosmographic constraints. } \label{f2pwlaw}
\end{figure}

As first step let us consider the power-law model in the (Eq. (\ref{ftBengochea})): it turns out that the parameters $\alpha$ and $b$ can  be expressed as function of  $H_0$ and $\Omega_m$,  just requiring that they verify the Eq.  (\ref{f0T}) and $f'(T_0) = 1$. Varying $\Omega_ m$ in the range $\left(0.2,0.33\right)$ and $h$ in the range $\left(0.65, 0.7\right)$ it turns out that both $f_2$  and $f_3$  are in agreement with the  cosmographic constraints, showing a consistent overlap, as shown, for instance, in Fig. (\ref{f2pwlaw})

Moreover,  let us consider the model (\ref{ftmyrza}), and  impose that the Eqs.
(\ref{f0T}) and $f'(T_0) = 1$ are satisfied, thus providing $\alpha$\, and $\beta$ as function of $\delta$:
\begin{equation}
\alpha(\delta) = \frac{2-\Omega_{m0} - [1 + (\Omega_{m0} - 2) \delta] \ln{T_0}}{1 + (\delta - 1) \ln{T_0}} \ ,
\label{am}
\end{equation}
\begin{equation}
\beta(\delta) = \frac{( \Omega_{m0}-1) T_0^{1 - \delta}}{1 + (\delta - 1) \ln{T_0}} \ ,
\label{bm}
\end{equation}
The derivatives $f_{i}$, with $i={2,3,4}$, indeed,  turn out to be parametrized by $\delta$. We estimate $\delta$, by imposing that
\begin{equation}
f_2{\delta}=f_2^{obs.}\,,
\label{f2eq}
\end{equation}
 being $f_2^{obs.}$ the best fit value (see Table \ref{tab2}), and calculate the higher order derivatives. However, otherwise than in  (\cite{vincft})  we limit to $f_3$ for testing our $f(T)$ models, since $f_4$ and $f_5$ depend also on the cosmographic parameters $s_0$ and $l_0$, which we are considering practically  weak and badly constrained. Indeed, we  use the $2\sigma$ confidence range to test the agreement of $f_3$ with the cosmographic constraints.
It turns out that  substituting  $\alpha$ and $\beta$ according to the Eqs. (\ref{am}), and (\ref{bm}), the  second degree Eq. (\ref{f2eq}) admits two different roots, which provide two different $2\sigma$ confidence range for $f_3$: actually we have that
\begin{eqnarray}
&&f_3^{\delta_1}\in(-0.51\,, 0.04)\\
&&f_3^{\delta_2}\in(8.5\,,8.9)\\
\end{eqnarray}
It turns out that both these values are  inconsistent with the cosmographic estimates, and, therefore, this model is not favoured by the observations.
A better situation can faced  in the case of model (\ref{ftmyrzabis}); we usually impose that the Eqs.
(\ref{f0T}) and $f'(T_0) = 1$ are satisfied, obtaining the following parametrized forms for $\alpha$ and $\beta$:
\begin{eqnarray}
&&\alpha=\frac{n (\Omega_m-2)-1}{n-1}\,,\\
&&\beta=-\frac{6^{1-n} (\Omega_m-3) H_0^{2-2 n}}{n-1}\,.
\end{eqnarray}
The parameter $n$ is evaluated imposing that $f_2(n)=f_2^{obs.}$: it results that  $n\in (-0.008, 0.13)$,  so that $f_3^{n}\in(-0.51, 0.08)$. I t worth noting that in this case the these values are weakly inconsistent with the cosmographic constraints.
It is worth noting that a slightly different procedure has been applied in the case of the $\tanh$ $f(T)$ model described in Eq. (\ref{wuhongewi1}). Actually in this case the parameters $\alpha$ and $n$ can be expressed as function of $\Omega_m$,  just requiring that they verify the Eqs. that (\ref{f0T}) and $f'(T_0) = 1$ :

\begin{eqnarray}
&&\alpha=\coth (1) \left(\Omega _m-2\right) 6^{\frac{1}{2-\Omega _m}+1-2 \text{csch}(2)} H_0^{\frac{2 \left(\Omega _m-3\right)}{\Omega _m-2}-4
 \text{csch}(2)}\,,\\
&&n=\frac{1}{\Omega _m-2}+2 \text{csch}(2)\,.
\end{eqnarray}
Thus, we can evaluate  $n$ , imposing that $f_2(n)=f_2^{obs.}$, providing $n\in\left(-0.055\,, -0.03\right)$.
Finally, in the case of the last model  (Eq. (\ref{Bamba})) it turns out that the parameters $\alpha$ and $\beta$ can be obtained as function of $\Omega_m$ only:
 \begin{eqnarray}
&&\alpha=\coth (1) \left(\Omega _m-2\right) 6^{\frac{1}{2-\Omega _m}+1-2 \text{csch}(2)} H_0^{\frac{2 \left(\Omega _m-3\right)}{\Omega _m-2}-4
 \text{csch}(2)}\,,\\
&&n=\frac{1}{\Omega _m-2}+2 \text{csch}(2)\,.
\end{eqnarray}
Varying $\Omega_ m$ in the range $\left(0.27,0.33\right)$ it turns out that $f_2$  is in agreement with cosmographic constraints, and also  $f_3$ shows a marginal overlap.


\section{Conclusions}

In this paper we use a cosmographic approach  to extract informations about the dynamics of the accelerating Universe considering only minimal assumptions (isotropy and homogeneity of the space-time) without choosing a priory any dynamical model of the dark energy. Our high-redshift analysis is based on  Taylor series expansions of the cosmological distances, and is aimed to constraint the so called {\it cosmographic parameters} up to the fifth order, inducing indirect constraints on any gravity theory. In particular we are interested in the teleparallel gravity theory, and are focused on its application  to cosmology,  i.e. in evaluating  the present day values of the function  $f(T)$ and its derivatives without solving the dynamical equations.  Our analysis is based on the Union2.1 Type Ia Supernovae (SNIa) data set, the Hubble diagram constructed from Gamma Ray Bursts luminosity distance indicators, some measurements of the function $H(z)$ derived from the age of passively evolving galaxies and from the BAO,  and some gaussian priors on the distance from the Baryon Acoustic Oscillations (BAO), and finally the Hubble constant h. We explore the 5-D space of parameters using the Markov Chain Monte Carlo technique, and following a Bayesian approach: we indeed use the median values as maximum likelihood value, in order to give more importance to the sampling of the distribution function of the marginalized parameters, rather than use the best fit accordance criterium. In order to  relate the constraints on the cosmographic parameters to the derivatives of $ f(T )$ we use the map defined through the Eqs. (\ref{f0T}---\ref{f4T}), which we evaluate along the final thinned {\it cosmographic chain}. To this aim we adopt a conservative approach, by considering that only the deceleration parameter, $q_0$, and the jerk, $j_0$, are well constrained, so that we limit our constraints on $f^{'''}$, since $f^{iv}$ depends on $s_0$ and $l_0$ also.
Deriving from a model independent method, these constraints have to be satisfied by any $f(T)$ theory. Therefore, we can investigate a priori, and without solving the field equation, a specific model by comparing the theoretically with the observed values of the derivatives of $ f(T )$. As an example, we have considered some most common models
It turns out that the estimations of $f_2$ and $f_3$ parameters for the model proposed by Bengochea et al. and Wu et al. are in agreement with the cosmographic constrains, showing a good overlap. Instead the estimations of the parameter for the first model proposed by Myrzakulov are weakly inconstintent with the cosmographic parameters, while the second model is not favoured by observations. Finally, in the case of the last model, it turns out that $f_2$ is in agreement with cosmographic constraints and $f_3$ shows a marginal overlap.
It is worth noticing how the renewed interest in cosmography has now opened the way to an alternative way to explore dark energy scenarios in different  gravity theories, such as $f(T)$.

\section*{Acknowledgments}

EP acknowledges the support of INFN Sez. di Napoli  (I.S. QGSKY and TEONGRAV).

\end{document}